\documentclass[usegrafix]{adqgem}
\usepackage{picinpar}
\newcounter{scholno}
\newcommand{\cf}{cf.~}

\newcommand{\body}[1]{body $\mathcal{#1}$}
\renewcommand{\thescholno}{{\Roman{scholno}}}

\newcommand{\vectv}{\vect{v}}
\newcommand{\vectu}{\vect{u}}
\newcommand{\vectr}{\vect{r}}
\newcommand{\vects}{\vect{s}}
\newcommand{\vectp}{\vect{p}}
\newcommand{\vectq}{\vect{q}}
\newcommand{\vectf}{\vect{f}}
\newcommand{\drate}[1]{\Delta#1}
\newcommand{\trate}[1]{\textd#1/\textd t}

\newenvironment{commentary}
{%
  \refstepcounter{scholno}
  \setlength{\parindent}{0pt}
  \setlength{\parskip}{10pt}
  \newenvironment{scholium}{\begin{center}\begin{minipage}{5in}{\small{\it Article} \thescholno.}\small}
    {\end{minipage}\end{center}}
}{}

\begin{document}

\workno{ART}{2001}{09A}
\worker{A. I. A. Adewole}
\monthdate{September 2001}
\mailto{aiaa@adequest.ca}
\work{Newton On Absolute Space: A Commentary}
\shortwork{Newton On Absolute Space: A Commentary}
\makefront

\begin{abstract}
Newton seems to have stated a quantitative relationship between the position of a body in 
relative space and the position of the body in absolute space in the first scholium of his 
Principia. We show that if this suspected relationship is assumed to hold, it will dispel 
many errors and misrepresentations that have plagued Newton's ideas on absolute space since
their inauguration.
\end{abstract}

\begin{center}
\section{The Scholium}\label{SS_SCHO}
\end{center}
\begin{commentary}\label{c1}
\begin{scholium}
Hitherto I have laid down the definitions of such words as are less known, and explained the 
sense in which I would have them to be understood in the following discourse. I do not define time, 
space, place and motion, as being well known to all. Only I must observe, that the vulgar conceive 
those quantities under no other notions but from the relation they bear to sensible objects. And 
thence arise certain prejudices, for the removing of which, it will be convenient to distinguish 
them into absolute and relative, true and apparent, mathematical and common.
\end{scholium}
Newton starts by telling us that time, space, place and motion are hard to define with precision and 
are better understood from experience. He then states that however we choose to understand them, we ought 
to distinguish two meanings of these notions. The first meaning is not stated here but is stated later 
on~[\cf\ref{c3}]. The second meaning is said to arise from the relationships that 
objects bear to one another. He describes the first meaning as absolute, true and mathematical. He 
describes the second meaning as relative, apparent and common. We are told that we need to distinguish 
the two meanings in order to remove certain prejudices and in order not to be vulgar or superficial.

That Newton's absolute space is strictly mathematical and abstract~[\cf\ref{c15}] needs to be emphasized. 
Our senses do not permit us to perceive absolute space or absolute motion and this is true unconditionally. 
Whatever space and motion we perceive through our senses is by stipulation common and apparent. For
this reason alone, interstellar space cannot represent absolute space in any way~[\cf\ref{c3}]. 

Some people seem to think that by referring all motion to a single body or frame, this body can be 
regarded as absolute. This is strictly incorrect because, according to Newton~[\cf\ref{c8}], 
motion relative to a body will be absolute if and only if the body is truly and absolutely at rest. 
But the simple act of referring all motion to a single body does not place the body at absolute 
rest~[\cf\ref{c16}].

Other people seem to think that if, for example, we find an effect or a phenomenon that depends not 
only on the relative motion of two bodies but also on their individual motion relative to 
a third body, then this third body can be regarded as being at absolute rest. Again, this is 
incorrect: a laboratory cannot be regarded as absolute by arguing that Doppler effects depend on 
the individual motion of an observer and a light source relative to the laboratory.
\end{commentary}

\begin{commentary}\label{c2}
\begin{scholium}
Absolute, true, and mathematical time, of itself, and from its own nature flows equably without 
regard to anything external, and by another name is called duration: relative, apparent, and common time, 
is some sensible and external (whether accurate or unequable) measure of duration by the means of motion, 
which is commonly used instead of true time; such as an hour, a day, a month, a year.
\end{scholium}
Newton is describing the two meanings of time that he advocates~[\cf\ref{c1}]. Absolute, true and 
mathematical time is said to change without regard to anything else (such as space). We are told that 
the changes in absolute time occur by themselves because of the nature of absolute time, an assumption 
that allows us to assign a meaning to {\em changes} in all other things. We are told also that changes 
in absolute time are equable, an assumption that allows us to assign a meaning to {\em equable} 
changes in all other things.

In contrast to absolute time, relative and common time does not change or change equably from its own nature. 
Instead, changes and equable changes in common time are brought about by means of relative and common motion 
in order that our senses may perceive them. Examples are an hour, a day, a month, a year. 

There is no point in asking if absolute time as described by Newton really exists. We are 
to understand absolute time strictly as a mathematical abstraction, to be approximated in practice by 
relative time, for the purposes of describing motion~[\cf\ref{c10}]. It exists therefore in the same sense
that every other mathematical abstraction exists.
\end{commentary}

\begin{commentary}\label{c3}
\begin{scholium}
Absolute space, in its own nature, without regard to anything external, remains always similar and 
immovable. Relative space is some movable dimension or measure of the absolute spaces; which our senses 
determine by its position to bodies; and which is vulgarly taken for immovable space; such is the dimension 
of a subterraneaneous, an aereal, or celestial space, determined by its position in respect of the earth.
\end{scholium}

The two meanings of space that Newton advocates~[\cf\ref{c1}] are now being explained. It appears 
that Newton is not describing absolute and relative spaces strictly in these sentences. He seems 
instead to be describing volume elements, or what he calls the ``dimension'' or ``measure'', of the spaces. 

We are told that a volume element of absolute space is not movable. The obvious reason for this is that
if a volume element of absolute space were movable, then it would be necessary to introduce another 
space below absolute space, ad infinitum, and many of Newton's opponents may interprete this 
circumstance as a weakness in his arguments. To avoid this, he asserted that the motion of a 
volume element of absolute space is inconceivable.

In contrast, the motion of a volume element of relative space is conceivable, and we are warned not to 
vulgarly take this volume element as a volume element of absolute space. We are told also that celestial
or interstellar space is a relative space and not an absolute space as some people seem to think~[\cf\ref{c1}].

The statement that (a volume element of) absolute space is always similar can be interpreted to mean that 
bodies are equably distributed in this space. It can also be interpreted as giving additional emphasis to 
the immovability of the volume element of absolute space without saying anything about the distribution of 
bodies in the space. A third interpretation is that any quantity that depends on the absolute position of
a body should be a homogeneous function of this position.
\end{commentary}

\begin{commentary}\label{c4}
\begin{scholium}
Absolute and relative space, are the same in figure and magnitude; but they do not remain always 
numerically the same. For if the earth, for instance, moves, a space of our air, which relatively and 
in respect of the earth remains always the same, will at one time be one part of the absolute space 
into which the air passes; at another time it will be another part of the same, and so, absolutely 
understood, it will be perpetually mutable.
\end{scholium}

Newton seems to be saying that absolute and relative spaces are the same in dimensions (``figure'') 
and extent (``magnitude''), but that a volume element of one space is not numerically equal to a 
volume element of the other space. For example, he says, consider a volume of air that is always 
the same with respect to the earth. As the earth moves from one place to another, a volume 
element of the air will remain the same with respect to the earth. But a volume element of 
the air in absolute space will be constantly changing (``perpetually mutable''). 

This suggests that if we have a proper understanding of relative and absolute spaces, we may 
be able to deduce an expression for the volume element of relative space in terms of the 
volume element of absolute space and vice versa.
\end{commentary}

\begin{commentary}\label{c5}
\begin{scholium}
Place is a part of space which a body takes up, and is according to the space, either absolute 
or relative. I say, a part of space; not the situation nor the external surface of the body. For the 
places of equal solids are always equal; but their superfices, by reason of their dissimilar figures, 
are often unequal. 
\end{scholium}

Newton defines the place of a body as the volume (``part'') of space taken by the body, and
states that a body's place can be absolute or relative. The place of a body is not the 
body's situation or position, nor is it the body's surface area. The latter distinction is necessary 
because, he says, two bodies with dissimilar shapes may have unequal surface areas (or superficies) 
but have equal places (or volumes).
\end{commentary}

\begin{commentary}\label{c6}
\begin{scholium}
Positions properly have no quantity, nor are they so much the places themselves, as the properties 
of places. The motion of the whole is the same thing with the sum of the motions of the parts; that is, 
the translation of the whole, out of its place, is the same thing with the sum of the translations of 
the parts out of their places; and therefore the place of the whole is the same thing with the sum of 
the places of the parts, and for that reason, it is internal, and in the whole body.
\end{scholium}

Here we are being told that volume (``place'') is an additive quantity, in the sense that the volume of 
a composite body is equal to sum of the volumes of the bodies that constitute the composite body. 
The motion of the composite body is, similarly, the sum of the motion of the constituent bodies, 
which is to say that velocity (``motion'') is additive. 

We are asked to understand that position is a property of volume elements and should not be confused 
with the volume elements themselves. Also, that while a unique or proper value can be assigned to 
the volume of a body, the position of a body, whether relative or absolute, cannot be assigned a 
unique value or quantity. 
\end{commentary}

\begin{commentary}\label{c7}
\begin{scholium}
Absolute motion is the translation of a body from one absolute place into another; and relative 
motion, the translation from one relative place into another. Thus in a ship under sail, the relative 
place of a body is that part of the ship which the body possesses; or that part of its cavity which 
the body fills, and which therefore moves together with the ship: and relative rest is the continuance 
of the body in the same part of the ship, or of its cavity. But real, absolute rest, is the continuance 
of the body in the same part of that immovable space, in which the ship itself, its cavity, and all that 
it contains, is moved. 
\end{scholium}

Having stated that ``place'' refers to a part or volume of space, Newton now explains how a body moves 
in relative and absolute spaces by changing its place in these spaces. He defines what it means for 
a body to be at rest in the spaces. A body is at relative rest when its position or situation in relative 
space is constant, and a body is at absolute rest when its position or situation in absolute space is 
constant (i.e., not changing with time).

It should be noted here that since the notion of absolute rest presupposes the notion of absolute space, 
we cannot hope to discover absolute space by discovering a body at absolute rest. Instead, we must 
first recognize absolute space before we can possibly recognize whether or not a body is at absolute 
rest. In particular, referring all motion to a single body does not imply that the body is at 
absolute rest~[\cf\ref{c1}].
\end{commentary}

\begin{commentary}\label{c8}
\begin{scholium}
Wherefore if the earth is really at rest, the body, which relatively rests in the ship, will 
really and absolutely move with the same velocity which the ship has on the earth. But if the earth 
also moves, the true and absolute motion of the body will arise, partly from the true motion of the earth, 
in immovable space; partly from the relative motion of the ship on the earth; and if the body moves also 
relatively in the ship; its true motion will arise, partly from the true motion of the earth, in immovable 
space, and partly from the relative motions as well of the ship on the earth, as of the body in the ship; 
and from these relative motions will arise the relative motion of the body on the earth.
\end{scholium}

This is a clear statement by Newton that motion relative to a body (such as the earth) is absolute if and 
only if the body is really and absolutely at rest. Even if it is convenient or natural to use the earth 
as a reference for describing the motion of a body in the ship, this motion will not be the absolute motion 
of the body unless the earth itself is at absolute rest. And to show that the earth is at absolute rest, one 
must show that its absolute position is constant --- convenience or simplicity of referring all motion to 
the earth does not count~[\cf\ref{c20}].
\end{commentary}

\begin{commentary}\label{c9}
\begin{scholium}
As if that part of the earth, where the ship is, was truly moved toward the east, with a velocity 
of 10010 parts; while the ship itself, with fresh gale, and full sails, is carried towards the west, 
with a velocity expressed by 10 of those parts; but a sailor walks in the ship towards the east, 
with 1 part of the said velocity; then the sailor will be moved truly in immovable space towards the east, 
with a velocity of 10001 parts, and relatively on the earth towards the west, with a velocity of 9 of 
those parts.
\end{scholium}

Suppose we have identified or recognized a body at absolute rest. Let the earth move relative to this 
body with a velocity of 10010 units towards the east; let a ship move relative to the earth with a 
velocity of 10 units towards the west; and let a sailor move relative to the ship with a velocity of 
1 units towards the east. Then the velocity of the sailor relative to the earth is 9 units towards the 
west while his velocity relative to the body at absolute rest is 10001 units towards the east. Note 
that this example {\em assumes} that we have recognized a body at absolute rest; the reason for this
assumption is that motion relative to absolute space itself is absurd~[\cf\ref{c16}].
\end{commentary}

\begin{commentary}\label{c10}
\begin{scholium}
Absolute time, in astronomy, is distinguished from relative, by the equation or correlation of 
the vulgar time. For the natural days are truly unequal, though they are commonly considered as equal 
and used for a measure of time; astronomers correct this inequality for their more accurate deducing 
of the celestial motions. It may be, that there is no such thing as an equable motion, whereby time 
may be accurately measured. All motions may be accelerated and retarded, but the true, or equable, 
progress of absolute time is liable to no change.
\end{scholium}

After telling us that astronomers make corrections for inequalities in the natural days to obtain the 
ephemeris time used in accurate studies of celestial motion, Newton states that we may never be able 
to measure true time to arbitrary accuracy because there may be no such thing as an equable motion. 
Yet we must assume that changes in true time are not liable to inequalities; for if they were, it 
will be necessary for us to introduce another time below absolute time, ad infinitum, in order for 
these inequalities to be quantifiable.
\end{commentary}

\begin{commentary}\label{c11}
\begin{scholium}
The duration or perseverance of the existence of things remains the same, whether the motions 
are swift or slow, or none at all: and therefore, it ought to be distinguished from what are only 
sensible measures thereof; and out of which we collect it, by means of the astronomical equation. 
The necessity of which equation, for determining the times of a phenomenon, is evinced as well from 
the experiments of the pendulum clock, as by eclipses of the satellites of Jupiter.
\end{scholium}

Here Newton states that the necessity of using true time for determining the duration of events is 
demostrated by experiments on the pendulum clock as well as by eclipses of Jupiter's satellites~---
a possible reference to the practice of fitting observational data to a presumed true time.
\end{commentary}
 
\begin{commentary}\label{c12}
\begin{scholium}
As the order of the parts of time is immutable, so also is the order of the parts of space. 
Suppose those parts to be moved out of their places, and they will be moved (if the expression may 
be allowed) out of themselves. For times and spaces are, as it were, the places as well of themselves 
as of all other things. 
\end{scholium}

Newton is setting the stage for telling us why the notion of motion cannot be applied to absolute 
time and space. The reason that motion is inapplicable to absolute time and space is that if it 
were applicable, then the parts that move will move out of themselves, which is absurd. The 
absurdity can be avoided by introducing additional spaces and times, ad infinitum, but since this 
may be perceived as a weakness in Newton's scheme, he cleverly ruled out this approach by declaring 
absolute time and space to be the {\em primary} places of things~[\cf\ref{c3}].
\end{commentary}

\begin{commentary}\label{c13}
\begin{scholium}
All things are placed in time as to order of succession; and in space as to order of situation. 
It is from their essence or nature that they are places; and that the primary places of things should 
be moveable, is absurd. These are therefore the absolute places; and translations out of those places, 
are the only absolute motions.
\end{scholium}

Newton states his case very clearly here. We are to understand absolute time as an arrangement of things 
or events in order of succession; absolute space, as an arrangement of things or objects in order of 
situation. Only by changing place in space and time do we have motion; hence, only things which can 
change place in space and time are movable. Since time and space are primary places, to apply the 
notion of motion to them is absurd~[\cf\ref{c12}]. 

Motion relative to absolute space is equally absurd, because if a body were to move relative to 
absolute space, then one might be led to conclude that absolute space moves also relative to 
the body. Instead, we must understand absolute space as a standard for defining rest, not for 
defining motion. 
To define the absolute motion of one body, we must find another body that is at 
rest in absolute space, and the motion of the one body with respect to the other body 
will then be absolute~[\cf\ref{c16}].
\end{commentary}

\begin{commentary}\label{c14}
\begin{scholium}
But because the parts of space cannot be seen, or distinguished from one another by our senses, 
therefore in their stead we use sensible measures of them. For from the positions and distances of things 
from any body considered as immovable, we define all places; and then with respect to such places, we 
estimate all motions, considering bodies as transferred from some of those places into others. 
\end{scholium}

Newton reiterates that absolute motions, being mathematical and abstract~[\cf\ref{c1}], cannot be 
perceived by our senses. He asserts that the relative motion of one body must be referred to another 
body, and that when we refer such motion to a designated body or frame, we are assuming implicitly 
that this designated body is at absolute rest (``considered as immovable''), and that 
this assumption may be valid or invalid.

The validity of the assumption is of no consequence if we are not interested in the 
possibility that the designated body 
may itself be in motion. If we are interested in this possibility, then we must introduce another space, 
which is the absolute space, in which the motion of the designated body can be investigated.

If we find that the position or situation of the designated body in absolute space is constant, then we 
may conclude that the body is truly and absolutely at rest. If we however find that the absolute 
position of the body is not constant, then we must conclude that the body is not truly or 
absolutely at rest. 

In either case, nothing prevents us from referring all motion to this body. That is, a body is 
not required to be at absolute rest before motion can be referred to it. If the body is at 
absolute rest, then any motion referred to it will be absolute; and if the body is not at absolute 
rest, then any motion referred to it will not be absolute.
\end{commentary}

\begin{commentary}\label{c15}
\begin{scholium}
And so, instead of absolute places and motions, we use relative ones; and that without any 
inconvenience in common affairs; but in philosophical disquisitions, we ought to abstract from our 
senses, and consider things themselves, distinct from what are only sensible measures of them. For it 
may be that there is no body really at rest, to which the places and motions of others may be referred.
\end{scholium}

Newton tells us that ordinarily, we can choose not to investigate the motion of a designated body 
to which all motion is referred. But in philosophical inquiries we ought not be so complacent; if 
necessary, we should even be prepared to abstract from our senses, and to consider things that our 
senses may not be able to perceive directly. The chief reason for this, he says, is that a body whose 
absolute position is constant may very well be nonexistent, in which case, regardless of whichever body 
we refer motion to, this motion will not be absolute.
\end{commentary}

\begin{commentary}\label{c16}
\begin{scholium}
But we may distinguish rest and motion, absolute and relative, one from the other by their properties, 
causes and effects. It is a property of rest, that bodies really at rest do rest in respect to one another. 
And therefore as it is possible, that in the remote regions of the fixed stars, or perhaps far beyond them, 
there may be some body absolutely at rest; but impossible to know, from the position of bodies to one 
another in our regions whether any of these do keep the same position to that remote body; it follows 
that absolute rest cannot be determined from the position of bodies in our regions.
\end{scholium}

Newton emphasizes here that in order for us to describe the absolute motion of a body, we must first 
find another body that is at absolute rest. The reason is that it is absurd for a body to move relative 
to absolute space itself~[\cf\ref{c14}].
If we find a body that we know to be at absolute rest, then every other body that is at relative rest 
with respect to the first body is also at absolute rest. 

Suppose there is a designated body at absolute 
rest, say, in the regions of the fixed stars or far beyond these regions. 
Then if we observe that Jupiter (for example) is at relative rest with respect to this body, then we 
must conclude that Jupiter is also at absolute rest, and that any motion referred to Jupiter will 
be absolute. 

But the regions of the fixed stars are so far away that it is practically impossible 
for us to verify that Jupiter is at relative rest with respect to the designated body. Not to mention 
the problem of verifying that the designated body is indeed at absolute rest.
\end{commentary}

\begin{commentary}\label{c17}
\begin{scholium}
It is a property of motion, that the parts, which retain given positions to their wholes, 
do partake of the motions of those wholes. For all the parts of revolving bodies endeavour to 
recede from the axis of motion; and the impetus of bodies moving forward, arises from the joint 
impetus of all the parts. Therefore, if surrounding bodies are moved, those that are relatively at 
rest within them, will partake of their motion. Upon which account, the true and absolute motion of 
a body cannot be determined by the translation of it from those which only seem to rest; for the 
external bodies ought not only to appear at rest, but to be really at rest. 
\end{scholium}

We are told that since the components of a composite body or system of bodies move as a group, 
it is impossible to determine the absolute motion of a body in the group by referring this motion 
to another body in the group, unless we can certify that this other body is indeed at absolute rest. 
Thus we cannot determine the absolute motion of a planet in the solar system by referring the motion 
to the sun unless we can also show that the sun is truly and absolutely at rest~[\cf\ref{c8}].
\end{commentary}

\begin{commentary}\label{c18}
\begin{scholium}
For otherwise, all included bodies, beside their translation from near the surrounding ones, 
partake likewise of their true motions; and though that translation were not made they would not be really 
at rest, but only seem to be so. For the surrounding bodies stand in the like relation to the surrounded as 
the exterior part of a whole does to the interior, or as the shell does to the kernel; but, if the shell 
moves, the kernel will also move, as being part of the whole, without any removal from near the shell.
\end{scholium}

The argument here is that if the sun is not at absolute rest, for example, then all the planets 
will take part in the absolute motion of the sun. Therefore, even if the planets were at relative 
rest with respect to the sun, they will still not be truly at rest because they will be 
taking part in the absolute motion of the sun~[\cf\ref{c22}].
\end{commentary}

\begin{commentary}\label{c19}
\begin{scholium}
A property, near akin to the preceding, is this, that if a place is moved, whatever is placed 
therein moves along with it; and therefore a body, which is moved from a place in motion, partakes 
also of the motion of its place. Upon which account, all motions, from places in motion, are no 
other than parts of entire and absolute motions; and every entire motion is composed of the motion 
of the body out of its first place, and the motion of this place out of its place; and so on, until 
we come to some immovable place, as in the before-mentioned example of the sailor. 
\end{scholium}

Just as the components of a composite body can move as a whole, a relative place and anything placed 
in it can also move as a whole. Hence if we refer the motion of \body{A} to another \body{B}
which is also moving, the relative motion of \body{A} with respect to \body{B} is only a part of the 
absolute motion of \body{A}; to get the entire absolute motion of \body{A}, we must take the 
absolute motion of \body{B} into account.

Suppose that a body moves with a system of bodies, and that this system moves with another system, 
and so on. To get the absolute motion of the body, we must take into account the motion of all systems 
of which the body is a part, until we come to the outermost system whose motion must then be referred 
to a body at absolute rest.

Another way of stating this is that if we refer the motion of all bodies in the universe to a 
designated body, then the motion of this designated body can be described only by introducing 
absolute space. Referring motion to a different designated body does not evade the argument 
but merely passes the problem on to the new designated body.
\end{commentary}

\begin{commentary}\label{c20}
\begin{scholium}
Wherefore, entire and absolute motions can be no otherwise determined than by immovable places; 
and for that reason I did before refer those absolute motions to immovable places, but relative ones 
to movable places. Now no other places are immovable but those that, from infinity to infinity, do 
all retain the same given position to one another; and upon this account must ever remain unmoved; 
and do thereby constitute immovable space.
\end{scholium}

Consequently, says Newton, we can only determine absolute motions by knowing a body at absolute rest, 
and relative motions by knowing a body at relative rest, which is the reason for associating relative 
motions with relative spaces and absolute motions with absolute spaces. No space is absolute or 
immovable unless, from end to end, volume elements of the space retain their position with respect to 
one another. Only when the volume elements of a space remain unmoved does the space constitute an 
immovable or absolute space. As stated elsewhere~[\cf\ref{c3}], Newton seems to prohibit the 
volume elements of absolute space from moving in order to avoid the infinite regression that will 
otherwise become necessary, a regression that may be perceived by his opponents as a weakness in 
his scheme~[\cf\ref{c12}].
\end{commentary}

\begin{commentary}\label{c21}
\begin{scholium}
The causes by which true and relative motions are distinguished, one from the other, are 
the forces impressed upon bodies to generate motion. True motion is neither generated nor altered, 
but by some force impressed upon the body moved; but relative motion may be generated or altered 
without any force impressed upon the body. For it is sufficient only to impress some force on other 
bodies with which the former is compared, that by their giving way, that relation may be changed, 
in which the relative rest or motion of this other body did consist. 
\end{scholium}

To understand the argument here, consider \body{A} and \body{B} which are at a certain distance 
from each other in relative space. If we impress a force on \body{A}, then the absolute 
motion of \body{A} will be altered.
But the absolute motion of \body{B} will remain unaltered, because no force is impressed upon it,
although its relative motion with respect to \body{A} will be altered, because its distance from 
\body{A} will be changing, unequably, on account of changes in the absolute motion of
\body{A}~[\cf\ref{c23}].

Assuming that we know what impressed forces are, this argument effectively stipulates that the 
effect of an impressed force is to alter (1) the absolute motion of the body 
upon which the force is impressed, and (2) the relative motion of any other body with 
respect to the first body.

It should be noted here that (i) the terms ``relative motion'' and ``absolute motion'' have now
acquired a very precise meaning which Newton consistently uses in the remainder of the
scholium; and (ii) Newton says nothing, anywhere in the remainder of the scholium, about changes 
in the {\em relative motion} of a body upon which a force is impressed.
\end{commentary}

\begin{commentary}\label{c22}
\begin{scholium}
Again, true motion suffers always some change from any force impressed upon the moving body; 
but relative motion does not necessarily undergo any change by such forces. For if the same forces 
are likewise impressed on those other bodies, with which the comparison is made, that the relative 
position may be preserved, then that condition will be preserved in which the relative motion consists. 
\end{scholium}

The absolute motion of \body{A} is always altered by impressing a force on \body{A}, but the 
relative motion of \body{A} with respect to another \body{B} is not necessarily altered by 
impressing a force on \body{A}. This is because one can impress an equal force on \body{B}
in such a way as to preserve its position relative to \body{A} at all times. 

Accoding to Newton, therefore, there is a sense in which two bodies at a fixed distance in 
relative space can be said to be moving. This happens when the relative position of the 
bodies is preserved by impressing equal forces on the bodies; we know that the bodies are moving 
in some sense, even though they are at relative rest with respect to one another, because 
impressed forces always alter the absolute motion of the body they are impressed upon. 

Hence, while two bodies at a fixed distance in relative space may seem to be at rest, we may 
not conclude that they are truly at rest: it is also possible that they are experiencing 
equal alterations in their absolute motions.

A good analogy of Newton's argument is the following. Suppose we want to determine the growth
(defined as the time derivative of the height) of a person called John. If we are able to
measure only differences in height, then we need to have another person, say James, whose
height can be compared with John's. But for our calculation to be meaningful, we must also
ensure that James' height remains fixed during our measurements. A vulgar or superficial way 
of ensuring this is to compare James' height with, say, Jane's. But the fact that James and
Jane have equal heights during our measurements does not imply that James' height is truly
fixed: it is also possible that both James and Jane are experiencing equal changes in their
heights. Newton's solution to this problem will become evident later~[\cf\ref{c33}].
\end{commentary}

\begin{commentary}\label{c23}
\begin{scholium}
And therefore any relative motion may be changed when the true motion remains unaltered, 
and the relative may be preserved when the true suffers some change. Upon which accounts, true motion 
does by no means consist in such relations.
\end{scholium}

\begin{figwindow}[0, l, {\unitlength1mm\input{FIG1.LP}}, 
{Changing the relative motion of \body{B} without changing its absolute motion.\label{FIG1}}]
Consider again \body{A} and \body{B} at some distance in relative space, and suppose that we impress 
a force upon \body{A}, \figref{FIG1}. Then since we have not impressed a force upon \body{B}, its absolute 
motion remains unaltered; but its relative motion with respect to \body{A} is changed on account of 
the change in the absolute motion of \body{A}~[\cf\ref{c21}]. Newton concluded that the relative 
motion of a body may be changed without changing the absolute motion of the body.
\end{figwindow}

\begin{figwindow}[0, l, {\unitlength1mm\input{FIG2.LP}}, 
{Changing the absolute motion of \body{B} without changing its relative motion.\label{FIG2}}]
If we now impress an equal force on \body{B} so as to preserve its position with respect to \body{A}
at all times, as shown in \figref{FIG2}, then the absolute motion of \body{B} will be altered 
on account of the force impressed upon it, but its relative motion with respect to \body{A} 
will be unaltered on account of the equality of the forces impressed on \body{A} and 
\body{B}~[\cf\ref{c22}]. Newton concluded that the absolute motion of a body may be changed 
without changing the relative motion of the body.

And since we can change one kind of motion without changing the other kind, Newton concluded also that 
the absolute motion of a body is not directly proportional to, nor linearly dependent on, the 
relative motion of the body with respect to another body. Or equivalently, that the position 
of a body in absolute space is not linearly dependent on the position of the body 
in relative space.
\end{figwindow}
\end{commentary}

\begin{commentary}\label{c24}
\begin{scholium}
The effects which distinguish absolute from relative motion are, the forces of receding 
from the axis of circular motion. For there are no such forces in a circular motion purely relative, 
but in a true and absolute circular motion, they are greater or less, according to the 
quantity of the motion. 
\end{scholium}

Having told us how we can cause absolute motion to change, Newton now proceeds to tell us the effect 
that a change in absolute motion has on the moving body. We are told that changes in absolute motion, 
such as that of a body moving in a circle in relative space, results in the existence of forces that 
tend to make the body recede from the axis of the circular motion. We are told also that these forces 
are proportional  to the ``quantity of motion'' (defined elsewhere by Newton)
while they are completely absent in a circular motion that is purely relative.

There is nothing peculiar about this result because according to Newton, \body{B} will be in 
``purely relative'' circular motion with respect to \body{A} if and only if a force is impressed 
upon \body{A} to cause it to move in a circle. Intuitively, therefore, \body{B} cannot experience 
a force attributable to changes in its relative motion with respect to \body{A}. 

But \body{A} experiences a force that causes changes in its absolute motion, by hypothesis. 
What Newton now adds to his scheme is that (1) \body{A} will experience another force whose 
effect on \body{A} is to make this body recede from the axis of motion; and that (2) this 
so-called centrifugal force is proportional to the quantity of motion of \body{A}.
\end{commentary}

\begin{commentary}\label{c25}
\begin{scholium}
If a vessel, hung by a long cord, is so often turned about that the cord is strongly twisted, 
then filled with water, and held at rest together with the water; after, by the sudden action of 
another force, it is whirled about the contrary way, and while the cord is untwisting itself, the 
vessel continues, for some time in this motion; the surface of the water will at first be plain, as 
before the vessel began to move: but the vessel, by gradually communicating its motion to the water, 
will make it begin sensibly to evolve, and recede by little and little from the middle, and ascend 
to the sides of the vessel, forming itself into a concave figure (as I have experienced), and the 
swifter the motion becomes, the higher will the water rise, till at last, performing its revolutions 
in the same times with the vessel, it becomes relatively at rest in it. 
\end{scholium}

This classic example by Newton demonstrates a phenomenon that involves a transition from the 
example in \figref{FIG1} to the example in \figref{FIG2}, where \body{A} is the vessel and 
\body{B} is the water~[\cf\ref{c23}].

In the beginning, a force impressed on the vessel causes changes in (1) the absolute motion of the 
vessel, and (2) the relative motion of the water with respect to the vessel; this corresponds
to \figref{FIG1}. 

Then as the vessel communicates its motion to the water, a force is impressed upon the water to cause 
changes in (1) the absolute motion of the water, and (2) the relative motion of the vessel with 
respect to the water.

Finally, the changes in the absolute motions of the water and the vessel become equal, at which point 
the vessel and the water are at relative rest with respect to one another; this corresponds to
\figref{FIG2}. 

But we know for a fact that the vessel and the water are clearly and visually not truly at rest.
This is a beautiful demonstration of Newton's statement that there is a sense in which bodies at 
rest in relative space can be said to be moving; that is, that such bodies need not be 
at rest in absolute space as they may simply be experiencing equal changes in their 
absolute motions~[\cf\ref{c22}].

With this example, Newton effectively illustrates the three claims he made earlier~[\cf\ref{c23}]
regarding the independence of relative motions and absolute motions.
\end{commentary}

\begin{commentary}\label{c26}
\begin{scholium}
This ascent of the water shows its endeavour to recede from the axis of its motion; 
and the true and absolute circular motion of the water, which is here directly contrary to 
the relative, discovers itself, and may be measured by this endeavour. At first, when the 
relative motion of the water in the vessel was greatest, it produced no endeavour to recede from 
the axis; the water showed no tendency to the circumference, nor any ascent towards the sides of 
the vessel, but remained of a plain surface, and therefore its true circular motion had not 
yet begun. 
\end{scholium}

Newton has stated that changes in the absolute circular motion of a body manifest in the existence 
of a centrifugal force whose effect on the body is to make the body recede from the axis of 
motion~[\cf\ref{c24}]. 
This effect is shown, he now says, by the tendency of the water to rise up the sides of the vessel. 

At the start of the experiment, the force impressed on the vessel causes changes in the
absolute motion of the vessel, as well as in the relative motion of the water with respect 
to the vessel. Since we observe that, at this time, the water shows no tendency to ascend to the 
sides of the vessel, we may conclude that there are no changes in the absolute motion 
of the water even though there are changes in the relative motion of the water with respect to 
the vessel~[\cf\ref{c25}].
\end{commentary}

\begin{commentary}\label{c27}
\begin{scholium}
But afterwards, when the relative motion of the water had decreased, the ascent thereof 
towards the sides of the vessel proved its endeavour to recede from the axis; and this endeavour 
showed the real circular motion of the water perpetually increasing, till it had acquired its 
greatest quantity, when the water rested relatively in the vessel. And therefore this endeavour, 
does not depend upon any translation of the water in respect of the ambient bodies, nor can true 
circular motion be defined by such translation. 
\end{scholium}

As the vessel communicates changes in its absolute motion to the water, the water experiences 
a force which causes the absolute motion of the water to increase, while
causing the relative motion of the vessel with respect to the water to decrease. 

The increase in the absolute motion of the water manifests in the tendency of the water to ascend 
to the sides of the vessel, and since we do observe this tendency, we may conclude rightly that 
the absolute motion of the water is indeed increasing.

Eventually, the changes in the absolute motion of the vessel and the water become equal, which 
indicates that the absolute motion of the water has reached its maximum, when the vessel 
and the water are relatively at rest.

What Newton demonstrates with this experiment is that two bodies which are relatively at rest 
need not be absolutely at rest because they may be experiencing equal changes in their absolute 
motions~[\cf\ref{c25}]. A corollary of this is that two bodies in relative motion must be 
experiencing unequal changes in their absolute motions; or, equivalently, that relative motion 
between two bodies arises from differences in the absolute motions of the bodies~[\cf\ref{c33}].
\end{commentary}

\begin{commentary}\label{c28}
\begin{scholium}
There is only one real circular motion of any one revolving body, corresponding to only 
one power of endeavouring to recede from its axis of motion, as its proper and adequate effect; 
but relative motions, in one and the same body, are innumerable, according to the various relations 
it bears to external bodies, and like other relations, are altogether destitute of any real effect, 
any otherwise than they may partake of that one only true motion.
\end{scholium}

Consider \body{A}, \body{B}, \body{C}, \body{D}. If we impress some forces on the first three
bodies, these forces will cause changes in (1) the absolute motion of the three bodies, and
(2) the relative motions of \body{D} with respect to the three bodies. Hence \body{D} will have 
different relative motions with respect to the other bodies depending on the forces impressed 
upon each of these bodies. 

But \body{D} has only one absolute motion, and this motion can be changed only by impressing a force 
on \body{D}. The effect of a change in the absolute motion of \body{D} is the appearance of a
centrifugal force that, if \body{D} is in a circular motion, will tend to make this body recede from 
the axis of the motion.

As long as forces are impressed on all other bodies except a designated body, the relative motions 
of the designated body will change, but these changes will not make the designated body experience 
the effect of a centrifugal force. This, says Newton, is what distinguishes the relative motion 
of a body from the absolute motion of the body~[\cf\ref{c34}].
\end{commentary}

\begin{commentary}\label{c29}
\begin{scholium}
And therefore in their system who suppose that our heavens, revolving below the sphere of the 
fixed stars, carry the planets along with them; the several parts of those heavens and the planets, 
which are indeed relatively at rest in their heavens, do yet really move. For they change their position 
one to another (which never happens to bodies truly at rest), and being carried together with their heavens, 
partake of their motions, and as parts of revolving wholes, endeavour to recede from the axis of their motions.
\end{scholium}

Newton argues here that bodies truly at rest cannot move relative to one another, and contrapositively, 
that bodies in relative motion cannot be at absolute rest~[\cf\ref{c27}]. 
Since the heavenly bodies change their position 
relative to one another, they cannot be at absolute rest. Changes in their absolute motions must also be 
unequal, for otherwise they will be at relative rest. Therefore, they must be experiencing the effect
of a centrifugal force that should tend to make them recede from the axes of their 
circular (``revolving'') motions. 
\end{commentary}

\begin{commentary}\label{c30}
\begin{scholium}
Wherefore relative quantities are not the quantities themselves, whose names they bear, but 
those sensible measures of them (either accurate or inaccurate), which are commonly used instead of 
the measured quantities themselves. And if the meaning of words is to be determined by their use, 
then by the names time, space, place and motion, their measures are properly to be understood; 
and the expression will be unusual, and purely mathematical, if the measured quantities themselves 
are meant. 
\end{scholium}

These are arguably the second most important sentences by Newton (his most important sentences are 
uttered shortly afterwards~[\cf\ref{c33}]). We are told that relative quantities are only measures, 
either accurate or inaccurate, of the corresponding absolute quantities. 

Then in one simple sentence, we are told to ignore everything we have been told regarding 
the distinction between relative and absolute quantities. Instead, we are asked to understand these 
things in their relative sense because the meaning of words should be determined by usage, and to 
understand these things in their absolute sense will be unusual and ``purely mathematical''. 

After going to great lengths to explain why relative quantities must be distinguished from absolute 
quantities, Newton now formally exorcises absolute quantities from his scheme without fanfare.
Whereas he would have said previously that an impressed force causes changes in the 
absolute motion of a body, he would now say that an impressed force causes changes in the relative 
motion of the body. Everything that was said previously for absolute motions must now be understood 
to hold for relative motions. 

But Newton insists that while the meaning words should be determined by the way we use words, and that 
while we are much better off using relative quantities in his scheme, we must nevertheless not use 
this as an excuse to confound relative and common quantities with absolute and true 
quantities~[\cf\ref{c31}].
\end{commentary}

\begin{commentary}\label{c31}
\begin{scholium}
Upon which account, they do strain the sacred writings, who there interpret those words 
for the measured quantities. Nor do those less defile the purity of mathematical and philosophical 
truths, who confound real quantities themselves with their relations and vulgar measures.
\end{scholium}

Having clarified his stance on the subject, Newton now chides those that interprete space and time 
in the ``sacred writings'' as referring to the absolute quantities instead of the relative 
quantities. He also chides those that fail to distinguish relative quantities from absolute 
quantities. The action of the latter people, he says, is a denial of mathematically and 
philosophically obvious statements or truths.
\end{commentary}

\begin{commentary}\label{c32}
\begin{scholium}
It is indeed a matter of great difficulty to discover, and effectually to distinguish, 
the true motion of particular bodies from the apparent; because the parts of that immovable space, 
in which those motions are performed, do by no means come under the observation of our senses. 
\end{scholium}

Newton reiterates the abstract and the mathematical character of absolute spaces and 
motions~[\cf\ref{c15}]. He emphasizes that we are unable to perceive these things because our 
senses are simply not conditioned to perceive them.
This does not mean that they are metaphysical in character --- 
we simply need to understand them in order to understand {\em why} our senses cannot perceive them.
\end{commentary}

\begin{commentary}\label{c33}
\begin{scholium}
Yet the thing is not altogether desperate; for we have some arguments to guide us, 
partly from the apparent motions, which are the differences of the true motions; 
partly from the forces, which are the causes and effects of the true motion. 
\end{scholium}

This is arguably the most important sentence uttered by Newton in the whole 
scholium~[\cf\ref{c30}]. Faced with the fact that our senses are not conditioned to perceive 
absolute spaces and motions, one may wonder how the existence of these things can be 
inferred from the relative spaces and motions that our senses do perceive. Here Newton 
tells us that the problem is by no means intractable, and he proceeded to give us 
two clues, one kinematic and one dynamic.

The kinematic clue is that relative motions arise from differences or changes in absolute 
motions~[\cf\ref{c27}]; or more precisely, that the relative motion of a \body{A} with 
respect to another 
\body{B} results from a change in the absolute motion of \body{B}. This clue tells us that 
if $\vectv$ is the relative velocity of \body{B} with respect to \body{A} (so that 
$-\vectv$ is the relative velocity of \body{A} with respect to \body{B}) and 
$\drate{\vectu}$ is a measure of changes in the absolute velocity $\vectu$ of \body{B}, 
then we may write $\vectv = \drate{\vectu}$. 
But if $\vectr$ is the relative position of \body{B} with respect to \body{A} and if 
$\vects$ is the absolute position of \body{B}, then $\vectv = \drate{\vectr}$ and 
$\vectu = \drate{\vects}$. Consequently, we have $\vectr = \drate{\vects} + const$. 

This means that we are to regard the relative position of a body as if it were
the time derivative ($\drate{}\equiv\trate{}$) of another quantity in absolute space; 
this quantity then represents the 
absolute position of the body in absolute space. {\em All} of Newton's statements on 
kinematics in absolute space can be understood on the basis of this quantitative relationship 
between the relative position and the absolute position of a body.

The dynamic clue given by Newton is that forces are the causes as well as the effects of 
absolute motions; or more precisely, that a change in the
absolute quantity of motion of \body{B} is 
caused by the action of an impressed force on \body{B}, and that the effect of this
change on \body{B} is the existence of a centrifugal force acting on \body{B}.
This clue tells us that changes in the absolute quantity of motion of a body are a measure 
of the forces impressed upon the body, which means that if the sum of these forces is denoted by 
$\vectf$, then we may write $\vectf=\drate{\vectp}$ where $\vectp=k\drate{\vects}$ 
is the quantity of motion and $k$ is the quantity of matter, both as defined by Newton, 
and $\vects$ is the absolute position of the body.

If we now exorcise absolute quantities from the scheme, as Newton recommended~[\cf\ref{c30}], 
we must restate our conclusion from the second clue in terms of the relative quantity of
motion of a body, or $\vectf=\drate{\vectp}$ with $\vectp=k\drate{\vectr}$, which is of course 
Newton's second law of motion.

Had Newton decided to incorporate absolute quantities into his scheme instead of summarily 
exorcising them as he does, the two clues he gave would have rendered his second law as
$\vectf=\drate{\vectp}$ with $\vectp=k\vectr$ instead of $\vectp=k\drate{\vectr}$. He would
then have had to make some necessary adjustments in his calculations in order for his
explanations of phenomena to remain consistent with observations. 
In particular, he would have had to adjust his first law to state that a body will continue 
to be in a state of {\em relative rest} unless compelled otherwise by impressed forces. This 
will then imply that a body moving uniformly in a straight line in relative space 
is experiencing a force (as Aristotle would say).

Newton may have decided not to follow this path, if at all he thought about it, for at least two 
good reasons. First, it will make
absolute space indispensable to his scheme, which is contrary to his wishes~[\cf\ref{c30}]. Second,
it will make uniform motion require that a force be applied, which is contrary to the prevailing
principle of Galileo that only a change in motion, but not motion itself, requires the application 
of a force. Perhaps Newton stated Galileo's
principle as his first law, even though he must have known that it is implied by his second law,
precisely in order to emphasize that in his scheme, the principle is strictly inductive and not 
to be regarded as a deductive consequence of his definitions. In this reading, the first law
declares general acceptance of the principle, while the second law defines a particular 
implementation thereof.
\end{commentary}

\begin{commentary}\label{c34}
\begin{scholium}
For instance, if two globes, kept at a given distance one from the other by means of 
a cord that connects them, were revolved about their common centre of gravity, we might, 
from the tension of the cord, discover the endeavour of the globes to recede from the axis 
of their motion, and from thence we might compute the quantity of their circular motions. 
And then if any equal forces should be impressed at once on the alternate faces of the globes 
to augment or diminish their circular motions, from the increase or decrease of the tension of 
the cord, we might infer the increment or decrement of their motions; and thence would be found 
on what faces those forces ought to be impressed, that the motions of the globes might be most 
augmented; that is, we might discover their hindermost faces, or those which, in the circular 
motion, do follow. But the faces which follow being known and consequently the opposite ones 
that precede, we should likewise know the determination of their motions. And thus we might 
find both the quantity and the determination of this circular motion, even in an immense vacuum, 
where there was nothing external or sensible with which the globes could be compared. 
\end{scholium}

Having explained his scheme as much as he could, Newton concludes with another example illustrating 
how forces are the causes and the effects of changes in the absolute motion of a body. 
He argues that the tension 
in the cord joining the two globes will depend on changes in the quantity of the absolute motion 
of the globes, and the changes can be measured, at least in principle, by measuring the tension 
in the cord.
\end{commentary}

\begin{commentary}\label{c35}
\begin{scholium}
But now, if in that space some remote bodies were placed that kept always a given 
position one to another, as the fixed stars do in our regions, we could not indeed determine 
from the relative translation of the globes among those bodies, whether the motion did belong 
to the globes or to the bodies. But if we observed the cord, and found that its tension was 
that very tension which the motions f the globes required, we might conclude the motion to be 
in the globes, and the bodies to be at rest; and then, lastly, from the translation of the globes 
among the bodies, we should find the determination of their motions. 
\end{scholium}

Newton says here that if there are some remote bodies in addition to the globes, there will be 
changes in the relative motion of these bodies with respect to the globes, and he agrees that it 
may seem difficult to ascertain whether the bodies are moving while the globes are at absolute rest, 
or whether the globes are moving while the bodies are at absolute rest.  

But we can determine what is actually moving, he says, by examining the tension in the cord; and 
if we find a tension that we would expect if the globes were moving, then we may conclude that the 
globes are indeed moving. This conclusion is valid because we know that changes in the {\em absolute
motion} of the remote bodies will not lead to any tension in the cord joining the 
globes~[\cf\ref{c24}].
\end{commentary}

\begin{commentary}\label{c36}
\begin{scholium}
But how we are to collect the true motions from their causes, effects, and apparent 
differences; and, vice versa, how from the motions, either true or apparent, we may come 
to the knowledge of their causes and effects, shall be explained more at large in the following 
treatise. For to this end it was that I composed it.
\end{scholium}

Newton concludes by telling us that his purpose is to demonstrate how we can deduce the absolute 
motions that our senses cannot perceive from the relative motions that our senses do perceive. 
He also wants to show us how to deduce the forces that are the causes and the effects of 
changes in absolute motions from a knowledge of those changes. But having 
exorcised absolute quantities from his scheme with the explanation that the meaning of words 
should be determined by their usage~[\cf\ref{c30}], it is no wonder that Newton says 
very little about absolute quantities in the rest of the Principia.
\end{commentary}

\begin{center}
\section{The Rationale}\label{SS_MORALE}
\end{center}
It appears that Newton's absolute space is an abstract, mathematical or ``phase'' space, no more
metaphysical than the hodographic space introduced by Hamilton into celestial 
mechanics, or any other phase space in modern physics for that matter. According to Hamilton,
we are to treat or regard the velocity of a body as if it were the ``position'' of the body
in hodographic space. According to Newton, we are to regard the position of a body as if it were
the time derivative of another quantity in absolute space; this quantity then represents the 
``position'' of the body in absolute space. Remarkably, Newton's absolute space bears exactly 
the same relationship to relative space that relative space itself bears to Hamilton's 
hodographic space. We can therefore understand the connection between absolute space and
relative space by studying the connection between relative space and hodographic space.
An interesting consequence of this circumstance is that if we relax Newton's assumption that
an absolute space be immovable, we see immediately that Newton's space is only one out of an
infinite number of absolute spaces.
Formally,  one may start with the space of the position vector $\vectr$ and define 
vectors $\vectq_n$ and $\vectp_m$, for $n, m = 1\ldots\infty$, by the equations 
\begin{equation*}
\vectq_n = \frac{\textd^n \vectr}{\textd t^n},\quad
\vectr = \frac{\textd^m\vectp_m}{\textd t^m} 
\end{equation*}
with the space of $\vectq_n$ called the hodographic space of order $n$, and the space of 
$\vectp_m$ called the absographic space of order $m$; so that, for examples, velocity space is a
first order hodographic space, acceleration space is a second order hodographic space, 
Newton's absolute space is a first order absographic space, etc. 

This generalization illustrates what Newton means when he describes his absolute space as
``purely mathematical'' and not ``under the observation of our senses''. What he seems
to have overlooked is that although we perceive motion only in relative space, other organisms 
(or even humans with the aid of some technological devices) might be able to perceive 
motion in the other spaces. An organism that perceives motion in first order hodographic space
will refer to what we call ``velocity'' as ``position''. Assuming that such an organism 
can show us some response when it perceives a moving object, it will behave as follow. 
If we are at relative rest with 
respect to the organism, it will show no response. If we start to move with some speed in 
any direction, it will show a response. If we speed up, slow down, turn left or turn right, 
it will show a response. But if we move with the same speed in a straight line, 
the organism will show no response because it will not perceive us to be moving: in its 
world or perspective, we will be at {\em hodographic rest} because what it perceives as 
our ``position'' is not changing. But we know for a fact that we are not ``truly'' 
or ``absolutely'' at {\em relative rest} with respect to the organism. This is basically
what Newton is describing in his scholium.

The fundamental problem addressed by Newton in the scholium is that our senses are
not capable of perceiving {\em position} itself but only {\em differences} in position.
This limitation of our senses implies that in order for us to be able to describe 
the motion of a \body{A} mathematically, we need
to have another \body{B} whose position can be compared with that of \body{A}. If we base
our description on calculus, then in order for the description to be 
meaningful, it is necessary to ensure that the position of \body{B} remains fixed 
throughout the experiment. A superficial and incorrect way to ensure this is to compare
the position of \body{B} with the position of another \body{C}. But the fact that
\body{B} and \body{C} are at the same distance throughout the experiment does not
imply that the position of \body{B} is fixed, because it is also possible that 
both \body{B} and \body{C} are comoving. Most of Newton's statements pertain to an 
analysis of this problem in all its nuances. Newton's insight is to recognize
the need for introducing another space in which position itself becomes an 
observable quantity derived from differences in a more fundamental but 
unobservable quantity called ``absolute position''. Using this new space, the
requirement that the position of a body should be fixed means that the absolute 
position of this body should not change unequably. In particular, the requirement 
that the position of a body be {\em fixed at zero} means, in Newton's parlance, 
that the body be {\em absolutely at rest}.

It is sometimes argued that absolute space cannot exist unless it affects our 
observations. That this argument is absurd is shown by the fact that one
can apply it equally well to any other phase space in modern physics in general, 
and to first order hodographic space in particular. The existence of absolute space presently
has a lot more to do with understanding than with observations, because if we do not
understand absolute space enough to be able to give a quantitative expression
for the absolute position of a body, then of course it cannot possibly affect 
our observations. For example, we may not argue that physical phenomena should be
independent of the hodographic position of a body while at the same time
permitting these phenomena to depend on the velocity of a body. But the 
contradiction in the argument will become evident only after we realize that the
velocity of a body is precisely the hodographic position of the body. Similarly,
arguments regarding the independence of physical phenomena on absolute positions
are meaningful only after we have a quantitative expression for the absolute
position of a body. Only then can we assert this independence by requiring that no
quantity may be a function of this expression, and only after this can sensible attempts 
be made to evaluate the observational consequences of this requirement. 

To conclude this brief excursion into ``philosophical disquisitions'', it is useful
to reiterate that our freedom to refer motion to any designated body of our 
choice is not hindered by the existence of absolute space; or to put it differently, that
a body does not have to be at absolute rest before we can refer motion to it. We can
refer motion to any body or frame regardless of the state of the absolute motion of the
body. If the body happens to be at absolute rest, then any motion referred to it will
be absolute; and if the body happens to not be at absolute rest, then any motion
referred to it will not be absolute. As long as we are interested only in relative motions,
we can refer these motions to any body whatsoever. Only when we are interested in 
absolute motions do we need to find a body at absolute rest to which the motions 
can be referred. Hence the fact that {\em relative motions} can be referred to an
arbitrary body does not in any way prove, imply or constitute a mathematical or a physical 
statement of the nonexistence of absolute space. By the same token, our inability to distinguish
a state of rest from a state of uniform motion, which by the way holds with equal validity
in relative space {\em as well as in absolute space}, does 
not in any way prove or represent a physical statement of the nonexistence of 
absolute space. Evidence that Newton understood these things is quite clear,
because he introduced absolute space in spite of Galileo's principle,
and he was careful to ensure that everywhere the word ``rest'' appears in his laws, 
there also appears a clause pertaining to ``uniform motion''. One hopes that,
in due course, when certain powerful prejudices have been overcome, philosophers and physicists 
of all dispositions will come to understand these things as well as Newton did.

\vspace{10pt}\section*{Acknowledgement}
Many thanks to participants in the usenet discussions that motivated this paper;
particularly, to Tim Shuba (who provoked my interest in the subject), and to 
Dennis McCarthy (whose aether provided the inspiration).
The text of Newton's scholium used in the paper is copied and pasted from the webpage
{\tt http://members.tripod.com/\verb ~ gravitee/definitions.htm}.

\end{document}